\documentstyle[ltwol]{article}

\input{psfig}
\def\beq{\begin{equation}}
\def\eeq{\end{equation}}
\def\bea{\begin{eqnarray}}
\def\eea{\end{eqnarray}}
\def\bseq{\begin{subequations}}
\def\eseq{\end{subequations}}
\def\nn{\nonumber}
\def\dfrac{\displaystyle\frac}

\def\numt#1#2{#1 \times 10^{#2}}
\def\etal{{\it et al.}}

\def\eg{{\it e.g.~}}

\def\btiny{\begin{tiny}}
\def\etiny{\end{tiny}}
\def\bsc{\begin{scriptsize}}
\def\esc{\end{scriptsize}}
\def\bfoot{\begin{footnotesize}}
\def\efoot{\end{footnotesize}}
\def\bsm{\begin{small}}
\def\esm{\end{small}}
\def\bno{\begin{normalsize}}
\def\eno{\end{normalsize}}
\def\bla{\begin{large}}
\def\ela{\end{large}}
\def\bLa{\begin{Large}}
\def\eLa{\end{Large}}
\def\bLA{\begin{LARGE}}
\def\eLA{\end{LARGE}}
\def\bhu{\begin{huge}}
\def\ehu{\end{huge}}
\def\bHu{\begin{Huge}}
\def\eHu{\end{Huge}}
\def\bCe{\begin{center}}
\def\eCe{\end{center}}
\def\bFR{\begin{flushright}}
\def\eFR{\end{flushright}}
\def\bFL{\begin{flushleft}}
\def\eFL{\end{flushleft}}

\def\PR#1#2#3{Phys. Rev. {\bf #1}, #2 (#3)}
\def\PRL#1#2#3{Phys. Pev. Lett. {\bf #1}, #2 (#3)}
\def\PL#1#2#3{Phys. Lett. {\bf #1}, #2 (#3)}

\def\NP#1#2#3{Nucl. Phys. {\bf #1}, #2 (#3)}

\def\PTP#1#2#3{Prog. Theor. Phys. {\bf #1}, #2 (#3)}

\def\eqref#1{eq.(\ref{eqn:#1})}

\def\eqlab#1{\label{eqn:#1}}

\def\bmaT{\left(\begin{array}{ccc}}
\def\emaT{\end{array}\right)}
\def\bma{\left( \begin{array} }
\def\ema{\end{array} \right)}
\def\l{\left}
\def\r{\right}
\def\gsim{~{\rlap{\lower 3.5pt\hbox{$\mathchar\sim$}}\raise 1pt\hbox{$>$}}\,}
\def\lsim{~{\rlap{\lower 3.5pt\hbox{$\mathchar\sim$}}\raise 1pt\hbox{$<$}}\,}
\input psfig
\def\cU#1#2{{\tilde{U}_{#1}^{#2}}}
\begin{document}

\title{
Prospects of Very Long Base-Line Neutrino Oscillation Experiments\\
with the JAERI-KEK High Intensity Proton Accelerator
}

\vspace{1.5mm}

\author{
N. Okamura$^{1,a}$\\
and\\
M. Aoki$^{1,b}$,
K. Hagiwara$^{1,c}$, 
Y. Hayato$^{2,d}$,
T. Kobayashi$^{2,e}$,
T. Nakaya$^{3,f}$,
K. Nishikawa$^{3,g}$.
}
\address{
$^1${Theory Group, KEK, Tsukuba, Ibaraki 305-0801, Japan}\\
E-mail:~
$^a${naotoshi.okamura@kek.jp},
$^b${mayumi.aoki@kek.jp},
$^c${kaoru.hagiwara@kek.jp}\\}
\address{
$^2${
Inst. of Particle and Nuclear Studies, High Energy Accelerator
Research Org., KEK,\\ Tsukuba, Ibaraki 305-0801, Japan}\\
E-mail:~
$^d${hayato@neutrino.kek.jp},
$^e${takashi.kobayashi@kek.jp}\\
}
\address{
$^3${Department of Physics, Kyoto University, Kyoto 606-8502, Japan}\\
E-mail:~
$^f${nakaya@scphys.kyoto-u.ac.jp},
$^g${nishikaw@neutrino.kek.jp}
}

\twocolumn[\maketitle\abstracts{
  In this paper, we discuss physics potential of the Very Long
Base-Line (VLBL) Neutrino-Oscillation Experiments with the
High Intensity Proton Accelerator (HIPA), which is planned to be built
by 2006 in Tokaimura, Japan.
 We propose to use conventional narrow-band $\nu_\mu$ beams (NBB)
from HIPA for observing the $\nu_\mu \to \nu_e$ transition probability
and the $\nu_\mu$ survival probability.
 The pulsed NBB allows us to obtain useful information through
counting experiments at a huge water-Cherenkov detector
which may be placed in our neighbor countries.
 We study sensitivity of such an experiment to the neutrino
mass hierarchy, the mass-squared differences,
the mixing angles and the $CP$
phase of the $3\times 3$ lepton flavor mixing matrix (MNS matrix).
 The $CP$ phase can be measured with a 100kt detector
if both the mass-squared difference and
$U_{e3}$ elements of the MNS matrix are sufficiently large.
}]
\section{Introduction}
 Very long base-line (VLBL) neutrino-oscillation
experiments is one of the attractive experiments in the near future.
 In this paper, we discuss physics potential of the 
VLBL neutrino-oscillation experiments with the
High Intensity Proton Accelerator (HIPA),
which was approved by the Japanese government last December
and will be built by 2006 in Tokaimura,
Japan\footnote{{More information is on {\sf
http://jkj.tokai.jaeri.go.jp/}.}}.

 The JHF Neutrino Working Group proposes
the first phase neutrino-oscillation experiments with HIPA
and Super-Kamiokande, whose base-line length is 295 km.
 The Letter of Intent (LOI) is available
on their web site \cite{LOI}.
 In this paper, we discuss possible future  
VLBL neutrino-oscillation experiment
between HIPA and Beijing.
 The base-line length of this experiment is about 2100km.

 In our analysis, we propose to use the conventional 
pulsed narrow-band $\nu_\mu$
beams (NBB) from HIPA for observing the $\nu_\mu \to \nu_e$
transition probability and the $\nu_\mu$ survival probability
at a huge 100kt-level detector in Beijing.
 We can then obtain useful information through
counting experiments \eg by adopting a
water-Cherenkov detector.
 We study the sensitivity of such experiment to the neutrino
mass hierarchy, that is the sign of mass-squared differences,
the mixing angles and the $CP$ phase in the $3 \times 3$
lepton flavor mixing matrix,
the MNS (Maki-Nakagawa-Sakata) matrix \cite{MNS}, and
the magnitudes of the two mass-squared differences.
 We can determine the neutrino mass hierarchy pattern from
this experiment.
 The $CP$ phase can be measured if both the mass-squared difference
responsible for the solar neutrino oscillation
 and
all three mixing angles are sufficiently large.

\section{MNS matrix}
 In this analysis, we assume three
neutrino spices\footnote{{%
If the LSND \cite{LSND} results are confirmed, we have to revise
this analysis.}}.
We define the MNS matrix as
\begin{eqnarray}
 J^{\mu}_{cc} &=&
(\overline{d},\overline{s},\overline{b})
\gamma^{\mu}(1-\gamma_5)
{V_{_{\rm CKM}}^{\dagger}}
(u,c,t)^T \nonumber \\
&+&
(\overline{e},\overline{\mu},\overline{\tau})
\gamma^{\mu}(1-\gamma_5)
{V_{_{\rm MNS}}^{}}
(\nu_1^{},\nu_2^{},\nu_3^{})^T\,,
\end{eqnarray}
where $u,d,c,s,t,b$ are the quark mass-eigenstates,
$e,\mu,\tau$ are the charged-lepton mass-eigenstates, 
and $\nu_i$ ($i=1,2,3$) is the neutrino mass-eigenstate. 
The MNS matrix connects the mass eigenstate $\nu_i^{}$ ($i=1,2,3$)
to the flavor eigenstate $\nu_\alpha^{}$ ($\alpha = e, \mu,\tau$)
\begin{eqnarray}
 \nu_\alpha^{}=
\sum_{i=1}^3 \l({{V}_{_{\rm MNS}}}\r)_{{\alpha}{i}}^{}
~{\nu_i^{}}\,.
\end{eqnarray}
 ${V}_{_{\rm MNS}}$ can be parameterized as
\begin{eqnarray}
{V}_{_{\rm MNS}} = {U}_{_{\rm MNS}} {\cal P}
=
\bmaT
U_{e 1}    & U_{e 2}    & U_{e 3} \\
U_{\mu 1}  & U_{\mu 2}  & U_{\mu 3} \\
U_{\tau 1} & U_{\tau 2} & U_{\tau 3}  
\emaT
\bmaT
1 & 0 & 0 \\
0 & e^{i \varphi_2^{}} & 0 \\
0 & 0 & e^{i \varphi_3^{}} 
\emaT\,,
\end{eqnarray}
 where the matrix ${\cal P}$ is the Majorana-phase matrix.
 If neutrinos were not Majorana particles,
this phase matrix can be absorbed away by the 
phases of the right-handed neutrino fields.
 The neutrino oscillation experiments are not sensitive to
this phase matrix. 
 The matrix $U_{_{\rm MNS}}$ has 
three mixing angles and one $CP$ phase,
just like the CKM matrix.
 We can always take these upper-right elements,
 $U_{e2}^{}$, $U_{e3}^{}$, and $U_{\mu 3}^{}$,
as the independent parameters of the MNS matrix.
 Both $U_{e2}^{}$ and $U_{e3}^{}$
have the non-negative real values
and $U_{e3}^{}$ can be complex number,
which are related to the three mixing angles
and one $CP$ phase.
 The other elements are determined by the unitarity
conditions \cite{HO1}.

 We take into account existing constraints for the MNS matrix and 
the mass-squared differences as follows.
 From the atmospheric-neutrino oscillation measurements \cite{atm,atmT},
\begin{eqnarray}
\sin^22\theta_{_{\rm ATM}} = 1.0\,,
\delta m^2_{_{\rm ATM}} = \numt{3.5}{-3} \mbox{(eV$^2$)}\,.
\eqlab{3_1}
\end{eqnarray}
From solar-neutrino deficit observations \cite{sol,solT},
\begin{eqnarray}
\mbox{{large-mixing-angle MSW \cite{MSW} solution (LMA)~~~~~~~~~~~~~~}}\\
\sin^22\theta_{_{\rm SOL}} = 0.8\,,~~~
\delta m^2_{_{\rm SOL}} = \numt{5,15}{-5} \mbox{(eV$^2$)}\,,
\nn\\
\mbox{{small-mixing-angle MSW \cite{MSW} solution (SMA)~~~~~~~~~~~~~~}}\\
\sin^22\theta_{_{\rm SOL}} = \numt{7}{-3}\,,
\delta m^2_{_{\rm SOL}} = \numt{5}{-6} \mbox{(eV$^2$)}\,,\nn\\
\mbox{{vacuum oscillation solution \cite{VO}
(VO)~~~~~~~~~~~~~~~~~~~~~~~~}}\nn\\
\sin^22\theta_{_{\rm SOL}} = 0.9\,,
\delta m^2_{_{\rm SOL}} = \numt{7}{-11} \mbox{(eV$^2$)}\,.~~~~~~~
\eqlab{3_2}
\end{eqnarray}
From the
CHOOZ reactor experiments \cite{chooz},
\begin{eqnarray}
\sin^2 2\theta_{_{\rm CHOOZ}} < 0.1
\mbox{{ when }}
\delta m^2_{_{\rm CHOOZ}} > 10^{-3}\mbox{(eV$^2$)}.
\eqlab{3_3}
\end{eqnarray}

 The four independent parameters of the MNS matrix
are related to the above observables
and $CP$ phase :
\begin{eqnarray}
2{{\l|U_{e 3}\r|^2}} &=&
1 - \sqrt{ 1-\sin^2 2\theta_{_{\rm CHOOZ}}}\,,\\
2{{U_{\mu 3}^2}}&=&
1 -\sqrt{ 1-\sin^2 2\theta_{_{\rm ATM}}}\,,\\
2 {{U_{e 2}^2}} &=&
1 - { \l|U_{e3}\r|^2} -
\sqrt{\l(1 - { \l|U_{e3}\r|^2}\r)^2
-\sin^2 2\theta_{_{\rm SOL}}}\,, \nn\\
\\
arg\l({ U_{e 3}}\r) &=& -{\delta_{_{\rm MNS}}^{}}\,.
\end{eqnarray}
 The first three equations are obtained for the 
observed mass-squared differences
\begin{equation}
 \delta m^2_{_{\rm SOL}} = \l|\delta {m}^2_{12}\r| \ll
\l|\delta {m}^2_{13}\r|  = \delta m^2_{_{\rm ATM}}\,,
\end{equation}
where
$\delta m^2_{ij} \equiv m^2_j-m^2_i$.

\section{Probability and mass hierarchies}
The Hamiltonian in the matter is written as
\begin{equation}
{\cal H}_{\alpha\beta}
=
\dfrac{1}{2E_\nu}
\left(
 {\delta m^2_{13}} { U_{\alpha 3}^{}U_{\beta 3}^{\ast}}
+{\delta m^2_{12}} { U_{\alpha 2}^{}U_{\beta 2}^{\ast}}
+{{A} \delta_{\alpha e} \delta_{\beta e}}
\right)\,,
\eqlab{ham}
\end{equation}
where $\alpha$ and $\beta$ are flavor indices $(e,\mu,\tau)$
and $A$ measures the matter effect,
\begin{eqnarray}
 A&=&2\sqrt{2}G_F Y_e \rho {E_\nu} \nn\\
  &=& \numt{7.56}{-5}
\left(\dfrac{\rho}{\mbox{{g/cm$^3$}}}\right)
\left(\dfrac{E_\nu}{\mbox{{GeV}}}\right)\,.
\end{eqnarray}
 We assume that the matter density of the earth's
crust is constant at $\rho=3$.
 We can then diagonalize the Hamiltonian, \eqref{ham} as
\begin{eqnarray}
{{\cal H}} = \dfrac{1}{2E_\nu}
{\cU{}{}}
\bmaT
 {\lambda_1} & 0 & 0 \\
 0 & {\lambda_2} & 0 \\
 0 & 0 & {\lambda_3}
\emaT
{\cU{}{\dagger}}\,,
\eqlab{matt_U_m}
\end{eqnarray}
where $\cU{}{}$ is the MNS matrix in the matter.
We introduce the oscillation phase parameters 
in the matter as
\begin{equation}
 \tilde{\Delta}_{ij} = 
\dfrac{\lambda_j - \lambda_i}{\hbar c}
\dfrac{L}{2E}\,.
\eqlab{matt_mm}
\end{equation}

By using
\eqref{matt_U_m}
and
 \eqref{matt_mm}, 
the oscillation probability in the matter is obtained as
\begin{eqnarray}
P(\nu_\alpha \to \nu_\beta)
&=&
\delta_{\alpha\beta}
-4\left\{
{\rm Re} \left[
\cU{\alpha 1}{} \cU{\beta 1}{\ast} \cU{\beta 2}{} \cU{\alpha 2}{\ast}
\right]
\sin^2\frac{\tilde{\Delta}_{12}}{2}
\right.\nn\\
& &~~~~~+
{\rm Re}\left[
\cU{\alpha 2}{} \cU{\beta 2}{\ast} \cU{\beta 3}{} \cU{\alpha 3}{\ast}
\right]
\sin^2\frac{\tilde{\Delta}_{23}}{2}
\nn\\
& &~~~~~+
\left.
{\rm Re}\left[
\cU{\alpha 3}{} \cU{\beta 3}{\ast} \cU{\beta 1}{} \cU{\alpha 1}{\ast}
\right]
\sin^2 \frac{\tilde{\Delta}_{31}}{2}
\right\}
\nn \\
&&+2{\tilde{J}}
\left[
\sin {\tilde{\Delta}_{12}}+\sin {\tilde{\Delta}_{23}}+\sin 
{\tilde{\Delta}_{31}}
\right]\,,
\eqlab{prob1}
\end{eqnarray}
where
\begin{equation}
{\tilde{J}}=
{\rm Im}\left[
\cU{\alpha 1}{} \cU{\beta 1}{\ast} \cU{\beta 2}{} \cU{\alpha 2}{\ast}
\right]\,,
\end{equation}
for $(\alpha,\beta)=(e,\mu)\,,(\mu,\tau)\,,$ or $(\tau,e)$,
the Jarlskog parameter \cite{Jar} of the lepton sector in the matter.

 There are four types of mass hierarchies, as shown
at Table \ref{tab:hie1}.
\begin{table}[htbp]
\begin{center}
\vspace{-3ex}
\caption{Four type mass hierarchies.}\label{tab:hie1}
\begin{tabular}[t]{|c||c|c|c|c|}
\hline
case              & { I} & { II} & { III} & { IV} \\
\hline
$\delta m^2_{12}$ &  $\delta m^2_{_{\rm SOL}}$
                  &  -$\delta m^2_{_{\rm SOL}}$
                  &  $\delta m^2_{_{\rm SOL}}$
                  &  -$\delta m^2_{_{\rm SOL}}$\\
\hline
$\delta m^2_{13}$ &  $\delta m^2_{_{\rm ATM}}$
                  &  $\delta m^2_{_{\rm ATM}}$
                  &  -$\delta m^2_{_{\rm ATM}}$
                  &  -$\delta m^2_{_{\rm ATM}}$\\
\hline
\end{tabular}
\end{center}
\vspace{-2ex}
\end{table}
 When the MSW solutions are chosen
for the solar neutrino deficit problem,
 only the mass hierarchies  I and III
are relevant.
 Below we show results for all four mass hierarchies 
because the anti-neutrino oscillation probabilities
in the matter are related to those for neutrinos,
\begin{equation}
 P(\bar{\nu}_\alpha \to \bar{\nu}_\beta)_{
{\rm I, II, III, IV}}
=
 P({\nu_\alpha} \to {\nu_\beta})_{
{\rm IV, III, II, I}}\,,
\end{equation}
in the limit of spherically symmetric earth.
 The number of anti-neutrino events for 
the mass hierarchy case I
is obtained from that of neutrino
events for the hierarchy IV, by taking
account of the difference in flux and cross sections.

\section{Beam and Detector}
\begin{center}
\begin{figure}
\psfig{figure=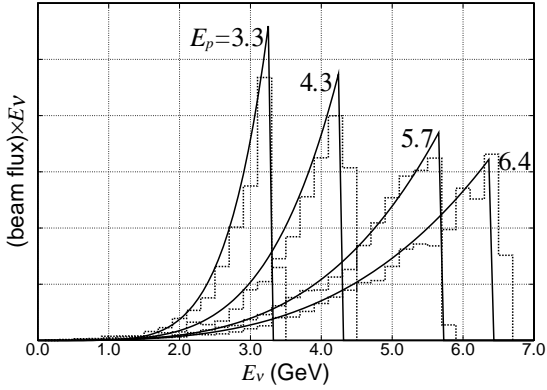,height=2.0in}
\caption{The Narrow Band Beam from HIPA.
The horizontal axis is the neutrino energy $(E_\nu^{})$
and the vertical axis is the beam flux times $E_\nu^{}$.
}
\label{fig:beamfit}
\end{figure}
\end{center}

 We examine pulsed-NBB,
whose shape is shown in Figure \ref{fig:beamfit}.
The horizontal axis gives the neutrino energy $E_\nu^{}$
and
the vertical axis gives the flux times $E_\nu^{}$.
 The histograms are 
obtained from the computer simulation and
we use the smooth fitted curves parameterized by
the neutrino peak energy, $E_p^{}$,
in the following analysis.
 We examine the case of 100kt water-Cherenkov detector.
 The detector should distinguish 
the e-like, $\mu$-like and neutral current events,
but we do not use the neutrino-energy information from 
the detector in this analysis.
That is, we assume the ``counting experiment'' in this analysis.

\section{Results}
The numbers of e-like and $\mu$-like events
are functions of only one variable, $E_p^{}$,
the peak energy of the NBB.
The number of events, $N(e(\mu):{E_p^{}})$
are obtained as
\begin{eqnarray}
N(e(\mu):{E_p^{}}) &=&
M N_A^{}
 {{\displaystyle \int}_{{0}}^{{E_p^{}}}}
 {\sigma^{^{cc}}_{e(\mu)}({E_\nu})}
 {\Phi({E_\nu},{E_p^{}})}\eqlab{event}\\
&& \hspace{15ex}\times
 {P_{\nu_\mu^{} \to \nu_{e(\mu)}^{}}({E_\nu})}
 {d}{E_\nu}\,,\nn
\end{eqnarray}
where
$M$ is the detector mass,
$N_{A}^{}$ is the Avogadro Number,
$\sigma^{^{cc}}_{e,\mu}(E_\nu^{})$ is 
the charged-current cross section
for each species,
$\Phi(E_\nu^{},E_p^{})$ is the NBB flux and
$P_{\nu_\mu^{} \to \nu_{e,\mu}^{}}({E_\nu^{}})$ is 
the transition or survival probability of $\nu_\mu$.

 First, we show the $CP$ phase dependence
of the expected number of e-like and $\mu$-like events
at Beijing for the NBB with 
$E_p^{}=$4GeV and 6GeV, and for 100ktyear.
 In this figure, we fix the mixing angles as
$\sin^22\theta_{_{\rm SOL}} = 0.8$,
$\sin^22\theta_{_{\rm CHOOZ}} = 0.1$, and
$\sin^22\theta_{_{\rm ATM}} = 1.0$,
and the mass-squared differences as
$\delta m^2_{_{\rm SOL}} = \numt{1.0}{-4} \mbox{(eV$^2$)}$
and
$\delta m^2_{_{\rm ATM}} = \numt{3.5}{-3} \mbox{(eV$^2$)}$.
\begin{figure}
\begin{center}
\psfig{figure=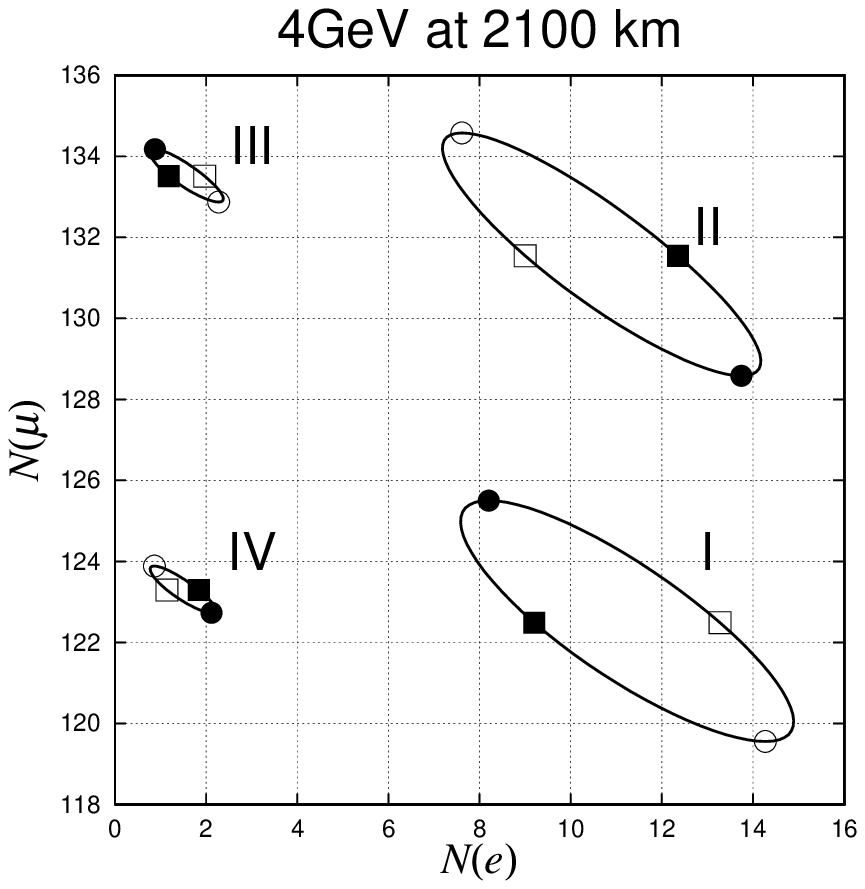,height=1.6in}
\noindent
\vspace*{-1.8in}
\hspace*{1.8in}
\psfig{figure=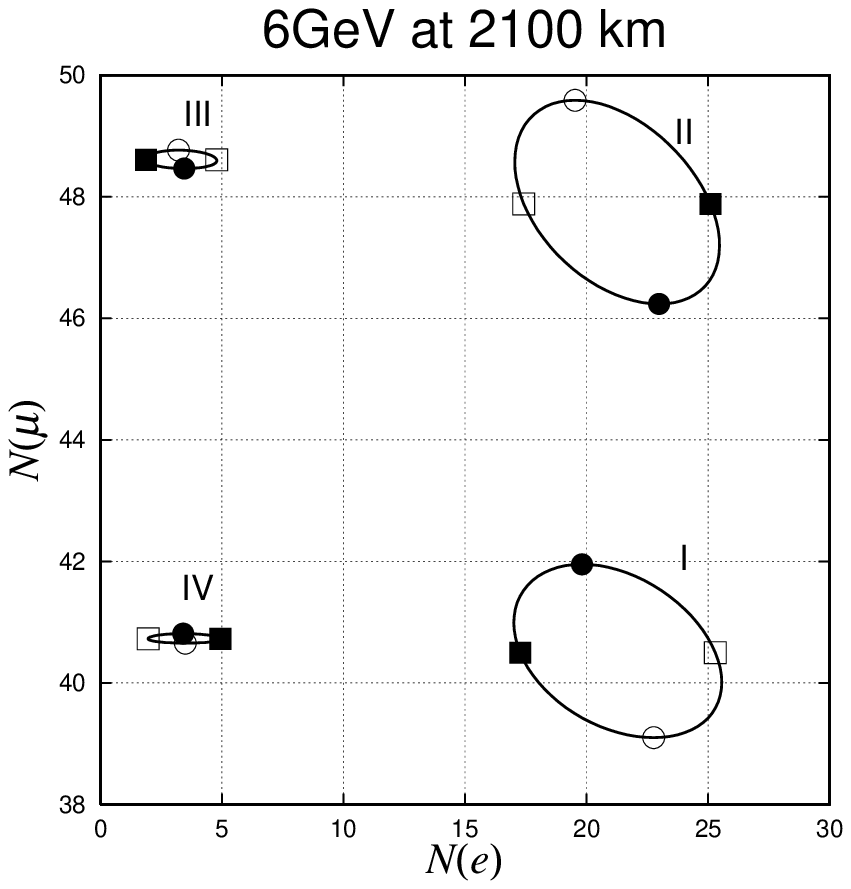,height=1.6in}
\end{center}
\caption{%
$CP$ phase dependence of $N(e)$ and $N(\mu)$
at Beijing for the NBB with $E_p^{}=4$GeV (left) 
and $E_p^{}=6$GeV (right).
$\delta_{_{\rm MNS}}^{} = 0^\circ$ (solid-circle),
$90^\circ$ (square),
$180^\circ$ (open-circle),
and $270^\circ$ (open-square).
 The results are shown for
$\sin^22\theta_{_{\rm SOL}} = 0.8$,
$\sin^22\theta_{_{\rm CHOOZ}} = 0.1$,
$\sin^22\theta_{_{\rm ATM}} = 1.0$,
$\delta m^2_{_{\rm SOL}} = \numt{1.0}{-4} \mbox{eV$^2$}$,
and
$\delta m^2_{_{\rm ATM}} = \numt{3.5}{-3} \mbox{eV$^2$}$.
}
\label{fig:CP_4_6GeV}
\end{figure}
 In each figure,
horizontal and vertical axis stand for the number of 
e-like and $\mu$-like events, respectively.
 The solid-circle, square, open-circle, and
open-square marks for $\delta_{_{\rm MNS}}^{} = 0^\circ$, 90$^\circ$,
180$^\circ$, and 270$^\circ$, respectively.

\begin{figure*}
\psfig{figure=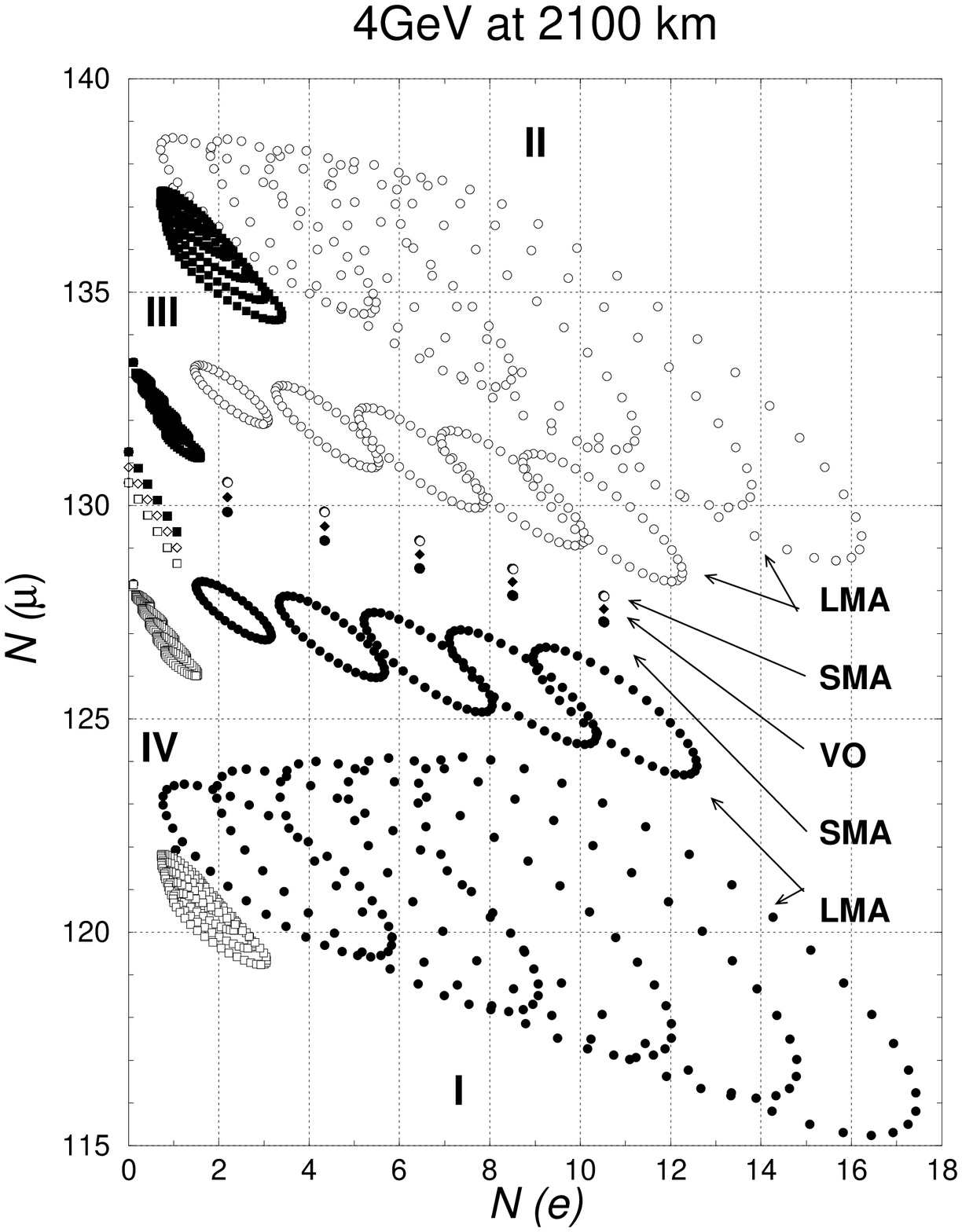,height=3.1in}
\vspace*{-3.1in}
\hspace*{4.0in}
\psfig{figure=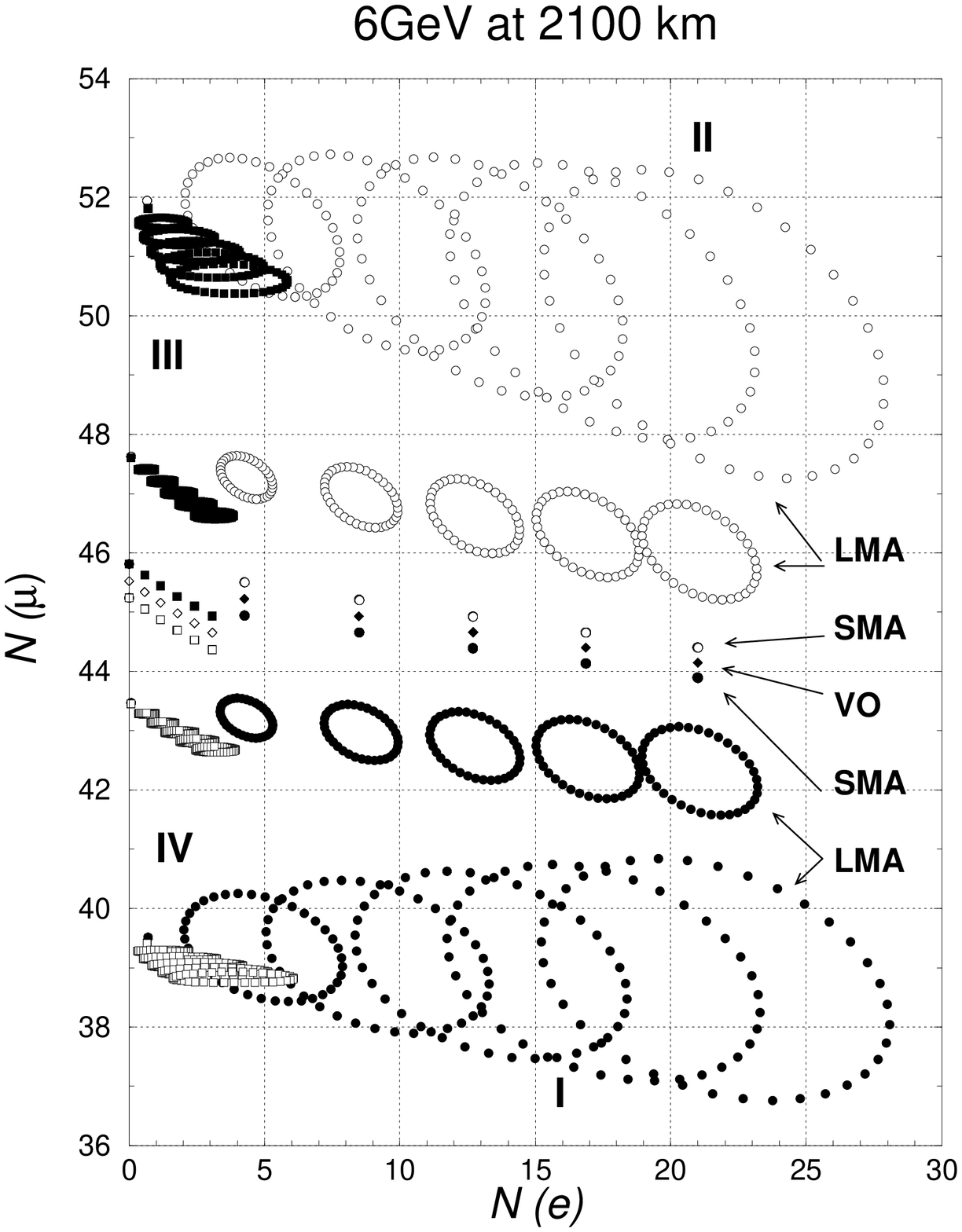,height=3.1in}
\caption{The neutrino parameter dependence of the
expected event numbers
for the NBB with $E_p^{}=4$GeV(left) and 6GeV(right)
and for 100ktyear at Beijing.
}
\label{fig:4_6GeV}
\end{figure*}

 In Figure \ref{fig:4_6GeV},
we show the expected numbers of e-like and $\mu$-like events
per year at a 100kt detector on Beijing
for the NBB with $E_p^{}=4$GeV (left) and 6GeV (right).
 Predictions of the VO scenario, the SMA solution, and
the LMA solution with $\delta m^2_{_{\rm SOL}}=\numt{5}{-5}$
and $\numt{15}{-5}$ eV$^2$ are shown.
 Five points or five circles with increasing $N(e)$ show expectations
for five values of $\sin^22\theta_{_{\rm CHOOZ}}$,
0.02, 0.04, 0.06, 0.08, to 0.1.
 The predictions of the VO scenario are common for the
mass hierarchies I and II, which are shown by the five dots
with larger $N(e)$,
and also for the case III and IV, the five dots with smaller $N(e)$.
The SMA scenario predicts smaller $N(\mu)$ in cases I and IV,
and larger $N(\mu)$ in cases II and III than that of the VO
scenario.
 The LMA scenario predicts even smaller $N(\mu)$ in cases I and IV,
and larger $N(\mu)$ for cases II and III,
where the deviation from the VO scenario
is more significant for larger $\delta m^2_{_{\rm SOL}}$.
\section{Summary}
 In this paper, we study the prospects of very long base-line
(VLBL) neutrino-oscillation experiments with the
High Intensity Proton Accelerator (HIPA).
 We present results for a
 VLBL experiment between HIPA and Beijing,
where the base-line length is about 2100 km.
 We propose to use 
 conventional pulsed-narrow-band $\nu_\mu$ beams
and a huge water-Cherenkov detector of 100kt in mass.
 The detector should distinguish e-like, $\mu$-like
and neutral current events but it is not required
to measure the neutrino-energy.
 We study the sensitivity of such an experiment to the
signs and the magnitudes of the neutrino mass-squared differences,
the mixing angles, and the $CP$ 
phase of the $3 \times 3$ lepton flavor mixing matrix (MNS matrix),
by using the $\nu_\mu \to \nu_e$ transition probability
and the $\nu_\mu$ survival probability.
 We find that the neutrino mass hierarchies can be determined from 
this experiment within several years.
 The $CP$ phase can be measured
if both the mass-squared difference  and
all the mixing angles of the MNS matrix are sufficiently large.



\section*{References}
\begin{thebibliography}{99}
\bibitem{LOI}
JHF Neutrino Working Group,
their web-site is {\sf  http://neutrino.kek.jp/jhfnu}.
%
\bibitem{MNS}
Z.Maki, M.Nakagawa and S.Sakata,
\PTP{28}{870}{1962}
%
\bibitem{LSND}
LSND Collaboration, \PRL{77}{3082}{1996}; \PRL{81}{1774}{1998}.
%
\bibitem{HO1} K. Hagiwara and N. Okamura, \NP{B548}{60}{1999}.
%
\bibitem{atm}
Kamiokande Collaboration, K.S.~Hirata {\it et al.},
 \PL{B205}{416}{1988};
{\it ibid.} {\bf B280}, 146 (1992);
Y.~Fukuda {\it et al.}, \PL{B335}{237}{1994};
IMB Collaboration, D.~Casper {\it et al.}, \PRL{66}{2561}{1991};
 R.~Becker-Szendy {\it et al.}, \PR{D46}{3720}{1992};
SOUDAN2 Collaboration, 
 T.~Kafka, Nucl.~Phys. (Proc.~Suppl.) {\bf B35}, 427 (1994);
 M.C.~Goodman, {\it ibid.} {\bf 38}, (1995) 337; 
 W.W.M.~Allison {\it et al.}, \PL{B391}{491}{1997};
 hep-ex/9901024.
Super-Kamiokande Collaboration,
\PRL{81}{1562}{1998};
\PRL{82}{2644}{1999};
hep-ex/9903047.
%
\bibitem{atmT}
 Super-Kamiokande Collaboration, T.~Toshito,
talk at the ICHEP 2000, 
 July 27 - August 2, 2000 at Osaka, Japan.
%
\bibitem{sol}
Homestake Collaboration, B.T.~Cleveland {\it et al.},
 Nucl.~Phys. (Proc.~Suppl.) {\bf B38}, 47 (1995);
 Ap. J. {\bf 496}, 505 (1998);
Kamiokande Collaboration,
Y.~Fukuda {\it et al.}, \PRL{77}{1683}{1996};
GALLEX Collaboration, W.~Hampel {\it et al.},
 \PL{B388}{384}{1996};
SAGE Collaboration, J.N.~Abdurashitov {\it et al.},
 \PRL{77}{4708}{1996};
Super-Kamiokande Collaboration, Y.~Fukuda {\it et al.},
\PRL{81}{1158}{1998}, \PRL{82}{1810}{1999}.

\bibitem{solT}
Super-Kamiokande Collaboration, Y.~Takeuchi,
talk at the ICHEP 2000,
 July 27 - August 2, 2000 at Osaka, Japan.

\bibitem{MSW}
L. Wolfenstein, \PR{D17}{2369}{1978};
S.P. Mikheyev and A.Yu. Smirnov, Yad. Fiz. {\bf 42}, 1441 (1985)
[Sov.J.Nucl.Phys.{\bf 42}, 913 (1986)];
Nuovo Cimento {\bf C9}, 17 (1986).
%
\bibitem{VO}
B.Pontecorvo, Zh.Eksp. Teor. Fiz. {\bf 53}, 1717 (1967);
S.M. Bilenky and B. Pontecorvo, Phys. Rep. {\bf 41}, 225 (1978);
V.Barger, R.J.N. Phillips and K. Whisnant, \PR{D24}{538}{1981};
S.L. Glashow and L.M. Krauss, \PL{190B}{199}{1987}.

\bibitem{chooz}
The CHOOZ Collaboration, \PL{B420}{397}{1998};
M.~Apollonio \etal, \PL{B466}{415}{1999}.

\bibitem{Jar}
C. Jarlskog,
\PRL{55}{1039}{1985} and Z. Phys. {\bf C29}, 491 (1985).
\end{thebibliography}
\end{document}